\documentclass[sigconf]{acmart}

\usepackage{booktabs} 

\usepackage{amsmath}
\usepackage{amssymb}
\usepackage{multirow}
\usepackage{graphicx}

\usepackage{gensymb}
\usepackage{tabularx}
\usepackage{tabu}

\usepackage{xcolor}

\usepackage{xcolor}
\newcommand\bcolorbox[2]{\bcbaux{#1}#2 \endbcb}
\def\bcbaux#1#2 #3\endbcb{%
  \colorbox{#1}{\strut#2}%
  \ifx\relax#3\relax\def\next{}\else%
    \colorbox{#1}{ \strut}%
    \allowbreak%
    \def\next{\bcbaux{#1}#3\endbcb}%
  \fi%
  \next%
}
\usepackage{amssymb}
\usepackage{graphicx}
\usepackage{booktabs}
\usepackage{url}
\usepackage{tabu}
\usepackage{xcolor}

\copyrightyear{2019}
\acmYear{2019} 
\setcopyright{iw3c2w3}
\acmConference[WWW '19 Companion]{Companion Proceedings of the 2019 World Wide Web Conference}{May 13--17, 2019}{San Francisco, CA, USA}
\acmBooktitle{Companion Proceedings of the 2019 World Wide Web Conference (WWW'19 Companion), May 13--17, 2019, San Francisco, CA, USA}
\acmPrice{}
\acmDOI{10.1145/3308560.3316736}
\acmISBN{978-1-4503-6675-5/19/05}

\usepackage{balance}

\begin{document}
\title{Neural Check-Worthiness Ranking with Weak Supervision: Finding Sentences for Fact-Checking}

\author{Casper Hansen, Christian Hansen, Stephen Alstrup, Jakob Grue Simonsen, Christina Lioma}
\affiliation{
  \institution{Department of Computer Science, University of Copenhagen}
}

\begin{abstract}
Automatic fact-checking systems detect misinformation, such as fake news, by (i) selecting \emph{check-worthy} sentences for fact-checking, (ii) gathering related information to the sentences, and (iii) inferring the factuality of the sentences. Most prior research on (i) uses hand-crafted features to select check-worthy sentences, and does not explicitly account for the recent finding that the top weighted terms in both check-worthy and non-check-worthy sentences are actually overlapping \cite{Le2016TowardsAT}.
Motivated by this, we present a neural check-worthiness sentence ranking model that represents each word in a sentence by \textit{both} its embedding (aiming to capture its semantics) and its syntactic dependencies (aiming to capture its role in modifying the semantics of other terms in the sentence). Our model is an end-to-end trainable neural network for check-worthiness ranking, which is trained on large amounts of unlabelled data through weak supervision. Thorough experimental evaluation against state of the art baselines, with and without weak supervision, shows our model to be superior at all times (+13\% in MAP and +28\% at various Precision cut-offs from the best baseline with statistical significance). Empirical analysis of the use of weak supervision, word embedding pretraining on domain-specific data, and the use of syntactic dependencies of our model reveals that check-worthy sentences contain notably more identical syntactic dependencies than non-check-worthy sentences.

\end{abstract}

\keywords{Fact checking; Check worthiness; Deep learning; Weak supervision.}

\maketitle

\section{Introduction}
The fast and wide spread of misinformation (as opposed to true information) on social media \cite{vosoughi2018spread,zubiaga2018detection}, and the increasing use of social media as a source of news\footnote{\url{https://www.reuters.com/article/us-usa-internet-socialmedia/two-thirds-of-american-adults-get-news-from-social-media-survey-idUSKCN1BJ2A8}} has turned ``fake news'' into an important societal problem on a scale that requires automated solutions. 
An automated fact-checking \cite{thorne2018automated} pipeline typically consists of three steps: (i) selecting \emph{check-worthy} sentences (i.e.\ sentences that contain check-worthy claims and should be fact-checked), (ii) gathering related information to those sentences, and (iii) using this related information to infer the factuality of each check-worthy sentence. Prior research has so far focused mainly on steps (ii) and (iii), for instance by generating claim-specific queries and querying search engines for relevant supporting information \cite{karadzhov2017fully}, focusing on specific sources such as Twitter \cite{ba2016vera}, or exploiting knowledge graphs from e.g.\ Wikipedia \cite{ciampaglia2015computational}. 
These approaches assume an input of check-worthy claims. 
Considerably less research has focused on detecting such check-worthy claims, that is, determining not whether a sentence is true or not, but whether a sentence contains a check-worthy claim (and should be fact-checked) or not. 

\begin{figure}
    \centering
    \includegraphics[width=0.75\linewidth]{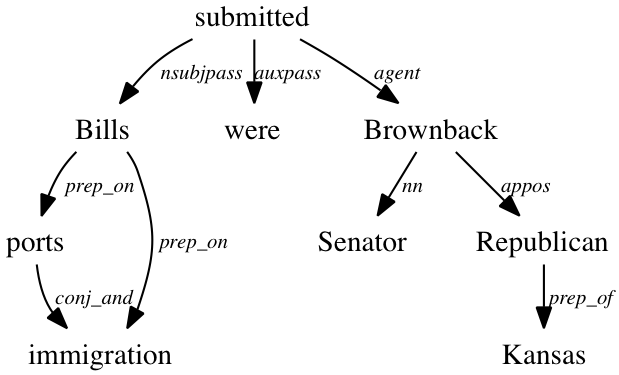}
    \vspace{-12pt}
    \caption{Syntactic dependencies example (from \cite{DBLP:conf/lrec/SilveiraDMBCBM14}).}
    \label{fig:syndep}
    \vspace{-10pt}
\end{figure}

Most research on check-worthiness detection uses hand-crafted features, such as bag-of-word representations, sentiment, and embedding averages \cite{cb,patwari2017tathya,jaradat2018claimrank,contextcb}. In addition, most work in this area does not explicitly account for the recent finding that the top weighted terms in both check-worthy and non-check-worthy sentences are actually overlapping \cite{Le2016TowardsAT}, hence compromising the effectiveness of bag-of-word based methods.



Motivated by the above, we present a neural check-worthiness sentence ranking model that uses a dual sentence representation: each word in a sentence is represented by \textit{both} its embedding (aiming to capture the semantics of that word from its context) and its syntactic dependencies (aiming to capture the role of that word in modifying the semantics of other words in the sentence, see Figure \ref{fig:syndep}). We train the network with these dual representations end-to-end. This allows to learn such descriptive features directly from the input data, rather than relying on predetermined hand-crafted features that may not be useful for the task, and hence to adapt the representations to the domain. To tackle the problem of having very little available training data, we use an existing check-worthiness system to weakly label sentences, and we use this weakly labelled dataset to pretrain our neural network. This is inspired by recent strong results \cite{nie2018multi,zamani2018sigir} in various information retrieval tasks with few labelled data, but large amounts of unlabelled data.

Thorough experimental evaluation against all state of the art baselines on political speeches from the 2016 U.S. election, shows our model to be superior in all comparisons (+13\% in MAP and up to +28\% at various Precision cut-offs from the best baseline, with statistical significance). We empirically trace this superior performance to the use of syntactic dependencies in the sentence representation, where we find check-worthy sentences to contain notably more identical syntactic dependencies than non-check-worthy sentences. 
Further analysis shows that the performance benefits of weak supervision increase with the amount of data used, and that embeddings trained on smaller domain-specific data, as opposed to general purpose embeddings trained on the larger Google News corpus, increase effectiveness. 
In addition, despite using deep learning (a family of models that is generally considered of low interpretability \cite{DBLP:journals/jzusc/ZhangZ18}), the attention weighting used by our model on a word level provides humanly interpretable output, where the parts of the sentence that are important for the check-worthiness prediction can be determined.

We \textbf{contribute} a competitive and interpretable end-to-end trainable neural network model for check-worthiness ranking, which uses a dual input representation of both word embeddings and syntactic dependencies. Weak supervision is used to pretrain the network on large amounts of unlabelled data.

\section{Related Work\label{sec:relatedwork}}


Given a sentence (also referred to as statement) as input, ClaimBuster \cite{cb,hassan2017claimbuster} outputs its check-worthiness score by extracting a set of features (sentiment, statement length, Part-of-Speech (POS) tags, named entities, and tf-idf weighted bag-of-words), and training a SVM classifier on these features to predict check-worthiness. Patwari et al.\ \cite{patwari2017tathya} present a system called TATHYA that is based on similar features, but that also includes as contextual features sentences immediately preceding and succeeding the one being assessed, as well as certain hand-crafted POS patterns. 
Gencheva et al.\ \cite{contextcb} also extend the feature set used by ClaimBuster to include more contextual features, such as the sentence's position in the debate text, and whether the debate opponent is mentioned. 
The work by Gencheva et al.\ has been extended to both English and Arabic \cite{jaradat2018claimrank}. In the recent CLEF 2018 competition on check-worthiness detection \cite{clef2018-total}, Zou et al. \cite{winnerclef2018} came first by using a large set of features (similarly to the above mentioned work), and doing feature selection with both a $\chi^2$-test and a linear SVM using a L1 regularizer. 

Prior work on neural networks for check-worthiness has been done by Konstantinovskiy et al.\ \cite{DBLP:journals/corr/abs-1809-08193}, who use InferSent \cite{conneau2017supervised} to derive a universal neural sentence representation and then train a logistic regression classifier on top of that. Similarly to our model, this approach also uses neural sentence embeddings. However, unlike our model, this approach uses \textit{universal} sentence representations, whereas we train our model \textit{end-to-end} to learn the representations directly from the input data, making the learning domain-specific. 

In the related domain of sentence factuality detection \citet{jimenez2018empirical} present a neural approach with multiple word embeddings. They artificially generate additional non-factual sentences to be used for training to increase robustness. Similarly to ours, this work also presents neural approaches that combine multiple word representations in order to improve performance. However, whereas the infusion of artificially generated non-factual sentences by \citet{jimenez2018empirical} allows weak supervision of a single class, we obtain weak labels independently of the type of sentence (i.e.\ not on a single class).



\section{Neural Check-Worthiness Sentence Ranking\label{sec:problemdef}}
Given a set of sentences as input, the aim is to rank them in descending order of check-worthiness. 
In order to better differentiate between sentences of varying degree of check-worthiness, We cast this as a ranking problem, as opposed to assigning a binary output to each sentence, following prior work \cite{cb,jaradat2018claimrank,contextcb} . Note that any ranked output can be made binary using an appropriate threshold, in case a subsequent fact-checking pipeline requires binary labelled sentences.
Given a set of sentences to be ranked, our model learns an end-to-end trained representation of each sentence. We describe first the representation of each word in the sentence (Section \ref{sec:word_rep_approach}), followed by the neural network architecture (Section \ref{sec:arch}).

\subsection{Neural sentence representation \label{sec:word_rep_approach} }
Our model uses a dual sentence representation: each word in a sentence is represented by \textit{both} its embedding and by its syntactic dependencies. The word embedding aims to capture the semantics of that word from its context. Embeddings of this type are generally well understood and have been found effective for check-worthiness detection \cite{DBLP:journals/corr/abs-1809-08193}. The syntactic dependencies of a word aim to capture the role of that word in modifying the semantics of other words in the sentence, for instance by being the subject or predicate of the sentence. We use a syntactic dependency parser \cite{choi2015depends} to map each word to its dependency (as a tag) in relation to the sentence structure, which we then represent as a one-hot-encoding. Dependency parsing is fast using state of the art tools (approximately 14,000 words per second) \cite{choi2015depends}.

Our motivation is that syntactic dependencies may be important for discriminating between common overlapping top-weighted words in both check-worthy and non-check-worthy sentences. The existence of common overlapping top weighted words in check-worthy and non-check-worthy sentences is reported by Le et al.\ \cite{Le2016TowardsAT} (see Figure 2 of \cite{Le2016TowardsAT} for examples), and to our knowledge, is not explicitly addressed by any prior check-worthiness approach. We posit that while these common top weighted words may not be distinguishable by their word representation, they may be distinguishable by their syntactic role in the sentence.



\subsection{Network architecture \label{sec:arch}}
Based on the above, each word in a sentence has two distinct encodings, that together represent the word.
We use this double representation of each word as input to a recurrent neural network (RNN) with GRU \cite{gru2014} memory units. The output from each word in the RNN is aggregated using an attention mechanism computed as $\alpha_t = \frac{\exp \left( \textrm{score} \left(h_t\right) \right) }{\sum_i \exp \left( \textrm{score} \left(h_i\right) \right)}$, where $h_t$ is the output of the GRU memory cell at time $t$, and $score(\cdot)$ is a learned function that maps the output to a scalar. The attention-weighted sum is fed to a fully connected layer, from which the output is predicted using a sigmoid activation function. 
We train the network using the RMSprop optimizer with binary cross entropy as the loss function (details in Section \ref{ss:tuning}).



\section{Experimental setup\label{sec:expsetup}}
We conduct two experiments: (I) we compare our model against state of the art baselines without weak supervision; (II) we use the ClaimBuster API (one of the baselines in experiment I) to weakly label a dataset of unlabelled political speeches (as described in Section \ref{sec:data}) and we use this weakly labelled data to pretrain the baselines and our model. ClaimBuster API is trained on a non-publicly available dataset and should therefore be considered as a black box baseline.

\subsection{Baselines}
We compare our model against these baselines (introduced in Section \ref{sec:relatedwork}), which have yielded state of the art performance at their date of publication: (1) ClaimBuster and its pretrained ClaimBuster API \cite{cb}, 
(2) TATHYA \cite{patwari2017tathya}, and the approaches by (3) Zou et al.\ \cite{winnerclef2018}, (4) Gencheva et al.\ \cite{contextcb}, and 
(5) Konstantinovskiy et al. \cite{DBLP:journals/corr/abs-1809-08193}. 

\subsection{Data\label{sec:data}}

\begin{table}
    \centering
    \scalebox{0.8}{
    \begin{tabular}{lccccc}
    \toprule
         Dataset & \#docs & \#sents. & sents. length & unique words & mean label \\ \hline
         Embed. tr. & 15,059 & 609,322 & 16.66 & 86,244 & - \\\hline
         Evaluation & 7 & 2,602 & 14.04 & 3,694 & 0.05 \\ \hline
         Weakly lab. & 161 & 37,732 & 16.53 & 13,314 & 0.24 \\ 
        \bottomrule
    \end{tabular}}
    \caption{Statistics of the embedding training, evaluation, and weakly labelled datasets.The evaluation dataset uses binary labels, and the weakly labelled dataset continuous labels in the interval $[0,1]$.}
    \label{tab:data}
    \vspace{-15pt}
\end{table}

\begin{figure}
    \centering
    \vspace{-7pt}
    \includegraphics[width=0.8\linewidth]{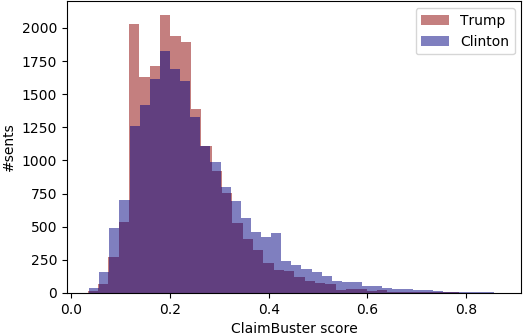}
    \vspace{-7pt}
    \caption{Histogram of the ClaimBuster scores used as the weak labels for each presidential candidate.}
    \label{fig:cbhist}
    \vspace{-13pt}
\end{figure}

We use three datasets for three different purposes: (1) the \textit{embedding training} dataset, used to train domain-specific embeddings for our model\footnote{None of the other approaches support training embeddings.}; (2) the \textit{evaluation} dataset, used to compare our model to the baselines without weak supervision; and (3) the \textit{weakly labelled} dataset, used to compare our model to the baselines with weak supervision. We describe these next (see Table \ref{tab:data} for statistics).

The \textbf{Embedding Training Dataset} contains documents related to \textit{all} U.S. elections available through the American Presidency Project\footnote{\url{https://web.archive.org/web/20170606011755/http://www.presidency.ucsb.edu/}}, e.g.\ press releases, statements, speeches, and public fundraisers, resulting in 15,059 documents. We use this dataset to pretrain a domain specific word embedding for our model (see Section \ref{sec:word_embedding_train}).

The \textbf{Evaluation Dataset} consists of a total of 2,602 sentences from 7 check-worthiness annotated political speeches from the 2016 U.S election. Out of these 7 speeches, 4 are by Donald Trump and are made available by the CLEF 2018 lab on automatic identification and verification of political claims \cite{clef2018-total}. The remaining are the inauguration and acceptance speech of Donald Trump and the acceptance speech of Hilary Clinton, and are made available by the authors of ClaimRank \cite{jaradat2018claimrank}. We choose the available PolitiFact annotated labels for this dataset. 

The \textbf{Weakly Labelled Dataset} consists of all publicly available speeches by Hillary Clinton and Donald Trump from the 2016 U.S. election, which are available through the American Presidency Project. This amounts to 37,732 sentences from 161 speeches not occurring in the evaluation dataset. We use the public API\footnote{https://idir-server2.uta.edu/claimbuster/} of ClaimBuster \cite{cb} to weakly label each sentence in all speeches. The ClaimBuster scores range from 0 to 1 (the higher the score, the more check-worthy the sentence), and are thus continuous as opposed to the binary labels from the evaluation dataset. The distribution of ClaimBuster scores can be seen in Figure \ref{fig:cbhist}, where we see that sentences by Hillary Clinton are overall labelled as slightly more check-worthy than those by Donald Trump.



\subsection{Tuning}
\label{ss:tuning}
We measure the effectiveness of the ranking outputted by our model and the baselines using mean average precision (MAP), and average precision at multiple cut-offs: P@5, P@10, P@20, and P@R, where R is the number of check-worthy sentences in a given test set. We optimize the MAP when tuning parameters.

We tune and evaluate the approaches using 7-fold cross validation, where the sentences from one speech (see Section \ref{sec:data}) act as testing data once, while sentences from the remaining speeches are used for training and validation. We use the sentences of each speech as input to the models. In all folds, we set aside 10\% of the training data for validation. Each fold-wise evaluation is repeated 5 times with randomly chosen validation data. We report the average score of each metric across the folds and repetitions. 

For ClaimBuster \cite{hassan2017claimbuster} and the model by Gencheva et al.\ \cite{contextcb}, we use the best performing setup described in \cite{contextcb}: a double layered fully connected neural network with layer sizes of 200 and 50 respectively. We validate these sizes by keeping the same ratio (4:1) between the layers and test the largest sizes of $\{50, 100, 200, 400\}$, test batch sizes of $\{64, 128, 256, 512\}$, and use a learning rate of 0.0001. For the approach by Zou et al.\ \cite{winnerclef2018} we use their multi-layer perceptron model with two hidden layers with sizes of 100 and 8 as per \cite{winnerclef2018}. We validate these sizes by keeping the same ratio (12.5:1) between layers and test the largest sizes of $\{50, 100, 200, 400\}$ as done earlier.
For TATHYA \cite{patwari2017tathya} we use the same multi-classifier approach and the same parameters as described in the original paper. For Konstantinovskiy et al. \cite{DBLP:journals/corr/abs-1809-08193} we use the same logistic regression approach as described in the original paper.

For our model, we evaluate the same layer sizes as above with a ratio of 4:1 between the number of neurons in the GRU cell and the single fully connected layer. We train the word embeddings using the word2vec skip-gram model \cite{word2vec} on the embedding training dataset of 15,059 domain specific documents described in Section \ref{sec:data}. We use standard parameters with a window size of 5 and sample 25 negative samples per word. 
For the syntactic dependencies of each word, we use the spaCy syntactic dependency parser \cite{choi2015depends}\footnote{The syntactic dependency parser is available at \url{https://spacy.io/}}. 

For the experiment with weak supervision, the ClaimBuster API returns a score from 0 to 1, indicating the degree of check-worthiness estimated by the system. In each fold in the 7-fold cross validation we find the threshold $\tau$ that splits the data and makes the fraction of check-worthy samples equal across the training without and with weakly labelled training data. Using these splits we evaluate two thresholding approaches: (1) Binarizing the labels based on $\tau$, and (2) truncating all values larger than $\tau$ to the value of $\tau$, and scaling the range $[0, \tau]$ into $[0,1]$. In the cross validation, our approach performs best with step (2) and the baselines perform best with step (1) (these are the settings we report in Section \ref{sec:results}). Note that step (2) can be considered as soft thresholding, as the labels are not binary. The weakly labelled data is used for pretraining the neural models or added to the training data for traditional supervised models.




\begin{table*}
    \centering
    \scalebox{0.9}{
    \begin{tabular}{@{}llllll|lllll@{}}
    \toprule
      & \multicolumn{5}{c}{Without Weak Supervision} & \multicolumn{5}{c}{With Weak Supervision}\\ 
         & MAP & P@5 & P@10 & P@20 & P@R $\;$ & MAP & P@5 & P@10 & P@20 & P@R \\ \hline
        ClaimBuster API\footnote{\url{http://idir-server2.uta.edu/claimbuster/}}\rule{0pt}{3ex} & 0.230 & 0.219 & 0.176 & 0.159 & 0.138 & - & - & - & - & -  \\ 
        TATHYA \cite{patwari2017tathya} & 0.136 & 0.072 & 0.062 & 0.074 & 0.039 & 0.147 & 0.061 & 0.047 & 0.060 & 0.043 \\ 
        ClaimBuster \cite{hassan2017claimbuster} & 0.176 & 0.170 & 0.112 & 0.105 & 0.078 & 0.224 & 0.198 & 0.147 & 0.138 & 0.121 \\ 
        
        Zou et al. \cite{winnerclef2018} & 0.187 & 0.143 & 0.105 & 0.099 & 0.086
        & 0.212 & 0.171 & 0.111 & 0.121 & 0.097\\
        
        Gencheva et al. \cite{contextcb} & 0.238 & 0.276 & 0.170 & 0.153 & 0.123 & 0.236 & 0.222 & 0.143 & 0.138 & 0.113 \\ 
        Konstantinovskiy et al. \cite{DBLP:journals/corr/abs-1809-08193} \rule[-1.2ex]{0pt}{0pt} & 0.267 & \textbf{0.314} & 0.186 & 0.178 & 0.149 & 0.233 & 0.220 & 0.146 & 0.142 & 0.113 \\ \bottomrule
        
        
        
        Our model \rule{0pt}{3ex}\rule[-1.2ex]{0pt}{0pt} & \textbf{0.278} & 0.291 & \textbf{0.194} & \textbf{0.193} & \textbf{0.159} &
        \textbf{0.302}$^{\blacktriangle\triangle}$ & \textbf{0.344}$^{\blacktriangle}$ & \textbf{0.238}$^{\blacktriangle\triangle}$ & \textbf{0.218}$^{\blacktriangle\triangle}$ & \textbf{0.189}$^{\blacktriangle\triangle}$ \\ \bottomrule
        
    \end{tabular}}
    \caption{Effectiveness of sentence check-worthiness ranking without and with weak supervision. 
    $\blacktriangle$ marks statistically significant improvements with respect to the overall best baseline at the 0.05 level using a paired two tailed t-test. $\triangle$ marks statistically significant improvements with respect to the best overall approach without weak supervision at the 0.05 level using a paired two tailed t-test. Significance comparisons are done on the average metric-wise performance in each of the 5 repeated runs.
    }
    \label{tab:comparison}
    \vspace{-15pt}
\end{table*}

\section{Results \label{sec:results}}
Table \ref{tab:comparison} shows the results of the experimental comparison of our model to the baselines without and with weak supervision. 
We see that our model outperforms all baselines across all metrics (with improvements ranging from +9-28\%), except P@5 (only without weak supervision) where our model is the second best performing approach. Note that P@k is known to be unstable, especially at small values of k \cite{kelly2010effects,Buckley20003effects}. 
The best performing baseline is the approach of Konstantinovskiy et al. \cite{DBLP:journals/corr/abs-1809-08193}, the only other approach using neural embeddings. This points out the effectiveness of neural word embeddings for this task.

The difference in performance between ClaimBuster and the ClaimBuster API is due solely to the quality of the training data (it is otherwise the same approach) and illustrates the effect of training data quality upon model performance. 
The fact that our model still notably outperforms the ClaimBuster API baseline shows the benefit of an end-to-end learned representation as opposed to a feature engineered one, for this task.

Only ClaimBuster, the approach of Zou et al.\ \cite{winnerclef2018}, and our model obtain notable improvements from the weakly labelled data (ClaimBuster yields a performance similar to that of the ClaimBuster API). 
TATHYA \cite{patwari2017tathya}, and the approaches by Gencheva et al.\ \cite{contextcb} 
and Konstantinovskiy et al.\ \cite{DBLP:journals/corr/abs-1809-08193} do not benefit from weak supervision, most likely because feature-engineering is used as opposed to learning the representation.

\subsection{Syntactic dependency similarity between check-worthy sentences \label{sec:syn_verify}}
Our syntactic dependencies representations aim to discriminate between top weighted words that are common in check-worthy and non-check-worthy sentences based on the syntactic roles of these words (see Section \ref{sec:word_rep_approach}). 
To verify this we compute the average overlap of unique syntactic dependency tags between the following three types of randomly sampled sentence pairs: 1) Pairs of $n$ sampled check-worthy sentences, 2) pairs of $n$ sampled non-check-worthy sentences, and 3) mixed pairs of $n$ sampled check-worthy and $n$ sampled non-check-worthy-sentences. We set $n=10$ and repeat the computations 1000 times. Table \ref{tab:syntac_overlap} displays the resulting average overlaps and their standard deviation. We observe that check-worthy sentence pairs have the highest average overlap of syntactic dependencies, and non-check-worthy the lowest, but both have a similar standard deviation. This indicates that syntactic dependencies are more similar in check-worthy sentences than in non-check-worthy sentences, and as such constitute a good discriminator between check-worthy and non-check-worthy sentences that otherwise contain an overlap of common top-weighted terms. Note that the average overlap of 7 common syntactic dependencies in check-worthy sentences practically applies to approximately half of the words in those sentences (the average sentence length is 14.04 in that dataset -- see Table \ref{tab:data}). Mixed sentences (both check-worthy and non-check worthy) have an average overlap in between that of check-worthy and non-check-worthy sentences, but with a larger standard deviation, indicating that syntactic dependencies from this mixed group act as a mixed and less stable discriminating signal.

As an example of the overlap problem of common top-weighted terms, consider the check-worthy sentence "\textit{Since president Obama came into office another two million hispanic americans have fallen into poverty}" and non-check-worthy sentence "\textit{I'm running to be a president for all americans}". In these cases the words \textit{president} and \textit{americans} are both important to describe the content, but have different syntactic dependencies (\texttt{compound}/\texttt{attr} and \texttt{nsubj}/\texttt{pobj}, respectively).

\begin{table}
    \centering
    \scalebox{0.9}{
    \begin{tabular}{lcc}
    \toprule
         Sentence pairs & Average Overlap & Standard deviation \\ \hline
         Check-worthy & 7.00 & 1.03 \\ 
         Non-check-worthy & 4.74 & 1.08 \\ 
         Mixed & 5.64 & 2.87 \\
        \bottomrule
    \end{tabular}}
    \caption{Average overlap of syntactic dependency tags and its standard deviation between three types of sentence pairs.}
    \label{tab:syntac_overlap}
    \vspace{-20pt}
\end{table}

\subsection{Impact of pretrained word embeddings}
\label{sec:word_embedding_train}
Our model is the only approach in Table \ref{tab:comparison} to have word embeddings trained on a domain specific dataset. All other approaches either use no word embeddings (ClaimBuster \cite{cb} and TATHYA \cite{patwari2017tathya}), use global word embedding averages (Zou et al.\ \cite{winnerclef2018} and Gencheva et al.\ \cite{contextcb}), or use a universally trained sentence representation based on global embeddings (Konstantinovskiy et al.\ \cite{DBLP:journals/corr/abs-1809-08193}). To isolate the effect of these domain-specific trained embeddings upon the performance of our model, we run our method as described in Section \ref{ss:tuning} but vary the pretraining of the embeddings as follows: 1) using no embeddings at all; 2) using randomly initialized embeddings which are not pretrained; 3) using general purpose embeddings pretrained with word2vec on Google News\footnote{\url{https://code.google.com/archive/p/word2vec/}}; 4) using our pretrained domain specific embeddings as described in Section \ref{sec:data}.
Table \ref{tab:word_embedding_exp} shows the results when varying the embedding strategy. We see that domain specific embeddings (Politics) obtains large improvements -- compared to the general purpose embedding -- with improvements up to +12-34\%. The last row of Table \ref{tab:word_embedding_exp} shows the performance without the syntactic dependency parsing (i.e., only the word embedding), which highlights the large performance increase from the syntactic dependency parsing. As expected, no pretraining of the embeddings leads to much lower performance, though MAP is still comparable to most baselines, except for Konstantinovskiy et al. \cite{DBLP:journals/corr/abs-1809-08193}. Not using embeddings at all severely drops overall effectiveness. Collectively these findings show that performance benefits more from training embeddings on smaller, yet domain-specific, data than on much larger but general domain data.

\begin{table}
    \centering
    \scalebox{0.9}{
    \begin{tabular}{llllll}
    \toprule        
         Emb. pretrain & MAP & P@5 & P@10 & P@20 & P@R \\ \hline
	 No embed.	& 0.184 & 0.153 & 0.116 & 0.103 & 0.087 \\ 
         No pretraining & 0.237 & 0.230 & 0.156 & 0.148 & 0.130 \\
         Google News & 0.268 & 0.262 & 0.178 & 0.184 & 0.143 \\ \hline
         Politics & \textbf{0.302} & \textbf{0.344} & \textbf{0.238} & \textbf{0.218}& \textbf{0.189} \\
        Politics w/o syn. dep. & 0.285 & 0.290 & 0.209 & 0.202 & 0.167 \\
        
        \bottomrule
    \end{tabular}}
    \caption{Performance per type of embedding pretraining. The last row shows the performance without the syntactic dependency parsing.}
    \label{tab:word_embedding_exp}
    \vspace{-20pt}
\end{table}

\subsection{Effect of varying the amount of weakly labelled data}

\begin{figure}
    \centering
    \vspace{-10pt}
    \includegraphics[width=0.75\linewidth]{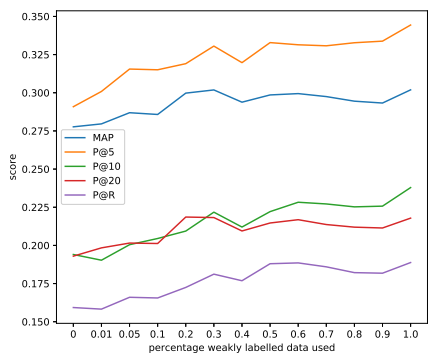}
    \vspace{-7pt}
    \caption{Impact of the amount of weakly labelled data upon our model (0 corresponds to no weakly labelled data).}
    \label{fig:our_weakExp}
    \vspace{-15pt}
\end{figure}

We analyse how our model, when used with weak supervision, scales with the amount of weakly labelled data, by reporting performance across the range of 0\% to 100\% weakly labelled data with 10\% increments. At each step we repeat the experiment 5 times with randomly sampled weakly labelled data, and report the average score. Figure \ref{fig:our_weakExp} displays performance as a function of the percentage of weakly labelled data. As expected, the scores of all metrics generally increase as the amount of data increases. However, the largest increases happen in the first 50\% of the data, and then small increases or stagnation for the remaining range up to around 90\%. The performance drop at 40\% may be explained by the limited number of repetitions of the sampling process, which was done due to runtime considerations. We expect that a larger number of repetitions would smooth out this slight drop.


\subsection{Model interpretability}
Check-worthiness detection can be part of semi-automatic or even fully manual fact checking processes, to filter claims that human fact checkers should investigate. In such cases, the output of check-worthiness detection should be easily interpretable by humans. Our model, despite being a deep learning model (generally considered to have low interpretability \cite{DBLP:journals/jzusc/ZhangZ18}) --
provides easily interpretable output to humans through the attention mechanism that is computed on a word level (see Section \ref{sec:arch}). This score can be used to highlight which parts of a sentence are important for the prediction of check-worthiness. Table \ref{tab:all_examples} illustrates this with a sample of true and false predictions made by our model. We see that sentences with high predicted scores (both correctly and incorrectly predicted as check-worthy) contain a quantifiable fact consisting of a relative large number, e.g.\ a large amount of money (\textit{trillion dollars, \$800 billion}), a high percentage (\textit{60 percent}), or a large collection of entities (\textit{nearly all other presidents combined}). Sentences with low predicted check-worthiness are more varied, but generally either lack a quantifiable element or are generally vague (\textit{buy American and hire American, fix the system, no patience for injustice}). We can also use the model to find seemingly incorrectly labeled sentences, as e.g. the non-check-worthy labelled sentences with high predictions could indeed be labelled as check-worthy instead, e.g. \textit{"our trade deficit in goods with world last year was nearly \$800 billion dollars"}.

\begin{table}[h]
    \centering
\scalebox{0.85}{

    \begin{tabu}{llX}
    \toprule
    $Y$ & $\Tilde{Y}$ & Sentence \\ \midrule
    
1 & 0.96 &  \fboxsep=0pt\relax\bcolorbox{red!13.77424807985601}{america }\bcolorbox{red!18.613882976368128}{has }\bcolorbox{red!18.62691719497064}{spent }\bcolorbox{red!27.550079569263094}{approximately }\bcolorbox{red!30.48249230927978}{six }\bcolorbox{red!51.97385212234011}{trillion }\bcolorbox{red!74.08506818039216}{dollars }\bcolorbox{red!76.52995950309338}{in }\bcolorbox{red!74.55881055509819}{the }\bcolorbox{red!77.61879940429213}{middle }\bcolorbox{red!81.96275537132487}{east }\bcolorbox{red!78.46529551050155}{, }\bcolorbox{red!75.21988695809091}{all }\bcolorbox{red!73.22952670200542}{this }\bcolorbox{red!77.21631591846284}{while }\bcolorbox{red!71.92857665458811}{our }\bcolorbox{red!76.01023286292074}{infrastructure }\bcolorbox{red!77.04716392956236}{at }\bcolorbox{red!79.63989513118554}{home }\bcolorbox{red!79.02970292780186}{is }\bcolorbox{red!86.46270913526283}{crumbling }\bcolorbox{red!87.31462322404359}{. }\\ \hline
1 & 0.95 &  \fboxsep=0pt\relax\bcolorbox{red!15.143014198735518}{today }\bcolorbox{red!13.970403103475064}{, }\bcolorbox{red!15.60298024256592}{our }\bcolorbox{red!25.069600598391986}{total }\bcolorbox{red!28.815444796527462}{business }\bcolorbox{red!37.03538760137079}{tax }\bcolorbox{red!59.059372549468435}{rate }\bcolorbox{red!61.71431318605343}{is }\bcolorbox{red!61.57092446551591}{60 }\bcolorbox{red!88.1576595947119}{percent }\bcolorbox{red!91.63127449322502}{higher }\bcolorbox{red!94.62979879245091}{than }\bcolorbox{red!94.83980178336512}{our }\bcolorbox{red!97.63880221208814}{average }\bcolorbox{red!99.14376159452581}{foreign }\bcolorbox{red!98.12131283097658}{competitor }\bcolorbox{red!98.04153743289964}{in }\bcolorbox{red!97.36368926265118}{the }\bcolorbox{red!95.87031516669408}{developed }\bcolorbox{red!96.19344702213118}{world }\bcolorbox{red!95.8217020733468}{. }\\ \hline
\midrule
1 & 0.26 &  \fboxsep=0pt\relax\bcolorbox{red!20.633940003827316}{its }\bcolorbox{red!26.317339812348212}{the }\bcolorbox{red!29.817115778266032}{same }\bcolorbox{red!36.43184173725808}{reason }\bcolorbox{red!32.61214367863078}{why }\bcolorbox{red!18.71560753150613}{she }\bcolorbox{red!9.105906788823312}{wo }\bcolorbox{red!6.067709512975636}{nt }\bcolorbox{red!7.408372434216742}{take }\bcolorbox{red!10.294388333941606}{responsibility }\bcolorbox{red!10.090562895826851}{for }\bcolorbox{red!5.136188867343888}{her }\bcolorbox{red!7.07363712198554}{central }\bcolorbox{red!7.80013289415685}{role }\bcolorbox{red!11.457411081266146}{in }\bcolorbox{red!9.008730839698877}{unleashing }\bcolorbox{red!7.2705048590806785}{isis }\bcolorbox{red!7.353864734325839}{all }\bcolorbox{red!8.630756789337024}{over }\bcolorbox{red!12.241172411562811}{the }\bcolorbox{red!22.9193499221711}{world }\bcolorbox{red!21.749427496606263}{. }\\ \hline

1 & 0.22 &  \fboxsep=0pt\relax\bcolorbox{red!23.742960433726235}{we }\bcolorbox{red!14.441256847699714}{will }\bcolorbox{red!12.968790891848347}{follow }\bcolorbox{red!14.271404286926762}{two }\bcolorbox{red!10.405375986943627}{simple }\bcolorbox{red!6.1111068402083735}{rules }\bcolorbox{red!3.5972986406971263}{; }\bcolorbox{red!6.798224797744524}{buy }\bcolorbox{red!8.069751780106726}{american }\bcolorbox{red!5.779988675205854}{and }\bcolorbox{red!11.155829683492344}{hire }\bcolorbox{red!14.513477155453623}{american }\bcolorbox{red!16.20180181916603}{. }\\ \hline


\midrule

0 & 0.04 &  \fboxsep=0.1pt\relax\bcolorbox{red!12.771916748341946}{millions }\bcolorbox{red!12.146585441638054}{of }\bcolorbox{red!13.460149962367236}{democrats }\bcolorbox{red!6.54210259488293}{will }\bcolorbox{red!3.150637686999202}{join }\bcolorbox{red!4.009241479951261}{our }\bcolorbox{red!5.320635339803268}{movement }\bcolorbox{red!3.091924914594089}{, }\bcolorbox{red!1.8394760218768413}{because }\bcolorbox{red!1.652879455838591}{we }\bcolorbox{red!1.2250931746754259}{are }\bcolorbox{red!0.6859613390794798}{going }\bcolorbox{red!0.4121214661220154}{to }\bcolorbox{red!0.21256528597084937}{fix }\bcolorbox{red!0.2541071765427194}{the }\bcolorbox{red!0.2562250841832458}{system }\bcolorbox{red!0.13096372867263797}{so }\bcolorbox{red!0.1405945934617469}{it }\bcolorbox{red!0.2115225402608933}{works }\bcolorbox{red!0.1758455670153186}{fairly }\bcolorbox{red!0.15411730870640736}{and }\bcolorbox{red!0.10365814913733752}{justly }\bcolorbox{red!0.12456366917757543}{for }\bcolorbox{red!0.0813939481758633}{all }\bcolorbox{red!0.14866993770014147}{americans }\bcolorbox{red!0.26005427595374087}{. }\\ \hline

0 & 0.05 &  \fboxsep=0.1pt\relax\bcolorbox{red!10.554704912617556}{i }\bcolorbox{red!8.19046514228829}{have }\bcolorbox{red!3.828836907112125}{no }\bcolorbox{red!3.745457492348365}{patience }\bcolorbox{red!3.286995968825031}{for }\bcolorbox{red!2.779914098573447}{injustice }\bcolorbox{red!4.961218529275401}{. }\\ \hline

\midrule
0 & 0.94 &  \fboxsep=0.1pt\relax\bcolorbox{red!13.289378384795492}{in }\bcolorbox{red!10.22726633203578}{the }\bcolorbox{red!15.37908300372075}{last }\bcolorbox{red!23.184943774025772}{eight }\bcolorbox{red!28.91336640417519}{years }\bcolorbox{red!25.778251415511154}{, }\bcolorbox{red!22.798421634683162}{the }\bcolorbox{red!22.144245040587027}{past }\bcolorbox{red!28.293276573108574}{administration }\bcolorbox{red!38.43553518201269}{has }\bcolorbox{red!43.99843303166971}{put }\bcolorbox{red!47.78069271557292}{on }\bcolorbox{red!49.438336461135634}{more }\bcolorbox{red!49.44023490384379}{new }\bcolorbox{red!56.07165375557105}{debt }\bcolorbox{red!69.3896391595883}{than }\bcolorbox{red!81.22536406665535}{nearly }\bcolorbox{red!78.90574103378104}{all }\bcolorbox{red!81.04710528091039}{of }\bcolorbox{red!79.05421962069161}{the }\bcolorbox{red!84.16155238198655}{other }\bcolorbox{red!86.05178502322663}{presidents }\bcolorbox{red!87.36865564573759}{combined }\bcolorbox{red!89.50435324200062}{. }\\ \hline
0 & 0.95 &  \fboxsep=0.1pt\relax\bcolorbox{red!7.356077589348136}{our }\bcolorbox{red!5.80245637883506}{trade }\bcolorbox{red!10.742862787799409}{deficit }\bcolorbox{red!13.96551161295899}{in }\bcolorbox{red!19.082569414662906}{goods }\bcolorbox{red!16.46208617980688}{with }\bcolorbox{red!17.83860776815882}{the }\bcolorbox{red!28.408349206678604}{world }\bcolorbox{red!39.572930231900585}{last }\bcolorbox{red!54.85439885990759}{year }\bcolorbox{red!43.92433874743001}{was }\bcolorbox{red!63.35195834909334}{nearly }\bcolorbox{red!84.91224887059771}{\$ }\bcolorbox{red!86.61384079317968}{800 }\bcolorbox{red!97.57805438586843}{billion }\bcolorbox{red!99.9999998086631}{dollars }\bcolorbox{red!98.73026419815201}{. }\\ \hline
    

    \end{tabu}}
    \caption{Check-worthy and non-check-worthy samples with high and low predictions ($\Tilde{Y}$) and ground truth labels ($Y$). Words are colored according to attention weight: the deeper the shade of red, the larger the attention score assigned.}
    \label{tab:all_examples}
    \vspace{-15pt}
\end{table}

\section{Conclusion\label{sec:conclusion}}
We have presented an end-to-end trainable neural network model for ranking check-worthy sentences. The model is pretrained via weak supervision from a large collection of unlabelled data and employs a recurrent neural network with a double representation of each word using domain specific word embeddings and the syntactic dependency parsing of a sentence. We evaluate our model on check-worthy annotated political speeches from the U.S. 2016 presidential election (following the same setting as in the official CLEF 2018 competition on check-worthiness detection \cite{clef2018-total} but using even more data). Our model does not make use of specialized hand-crafted features as most related work \cite{cb,patwari2017tathya,jaradat2018claimrank,contextcb}, but instead adapts the model and its representation to the domain by being trained in an end-to-end fashion. Thus, our model should by design be able to adapt to other check-worthiness tasks with results depending on the type of discourse and rhetoric. 
Our model effectively incorporates weak supervision: using an existing check-worthiness ranking system to weakly label political speeches significantly improved performance. Overall, our model outperforms all state of the art baselines in mean average precision and precision at different cut offs, with statistically significant +9-28\% gains from the best performing baseline. Further analysis revealed the significance of domain specific word embeddings, compared to traditional general purpose embeddings, and how check-worthy sentences share a syntactic similar structure than non-check-worthy sentences.

Future work consists of investigating further multiple weak signals and incorporating text discourse context into the model.

\begin{acks}
Partly funded by Innovationsfonden DK, DABAI (5153-00004A), and AMAOS (7076-00121B). 
\end{acks}

\bibliographystyle{ACM-Reference-Format}
\bibliography{main-wwwWorkshop}

\end{document}